# Theory of accelerated flows in a long wave approximation: High order shallow water equations (HSWE)

Arno Roland Ndengna Ngatcha[*]

*National Higher Polytechnic school of Douala, university of Douala, BP 2107, Douala, Cameroun.*

Corresponding author(*): arnongatcha@gmail.com.

**Abstract**

The majority of coastal flows are characterized by turbulence, rendering the application of shallow water equations an inadequate approach for their accurate description. This paper presents a theory for characterizing accelerated coastal flows, offering an enhanced representation of turbulent flows in the long-wave approximation. The fundamental premise of the theory is as follows: Fluctuating velocities present in the water flow are found to correlate and generate additional motion (or fluctuating motion), which is associated with a velocity of the same order and regularity as the mean flow velocity. The resulting high-order long-wave model is both stable and efficient, comprising classical shallow water equations and additional equations that describe the transport of kinetic energy due to turbulence. The model's detailed eigenstructure and Rankine-Hugoniot relations, steady state solutions are presented in this paper. The proposed theory shows promise for addressing a variety of engineering problems, including roll waves, hydraulic jumps, dam breaks, and flooding. The derived model can accurately address the problem of the mixing layer interaction with a free surface and its transition into a turbulent surface jet. Recent application to sediment transport gives interesting results and good numerical convergences.

## 1. Introduction

- **Shallow water based equations**

The Shallow water equations were first introduced by [1]. Shallow water models, also known as first-order long wave approximation models, are hyperbolic systems that have a wide range of applications, including dam break modeling, tsunami prediction, flooding, hydraulic jump, and sediment transport. These models provide a simpler approximation compared to the full three-dimensional Euler or Navier-Stokes equations with a free surface and two-phase models.

In the literature, a variety of first order shallow water based models has been developed or used over the past few century to address more complex problems including viscous effects [2], Coriolis forces [3], topography variations [4], [5], [6], density variations [7], [8], [9], earth's rotation [10], magnetodynamics [11], etc.

These models differ in terms of complexity, accuracy, inputs and outputs, as well as their spatial and temporal scales. Most theoretical methods used in the long wave theories are the result of deriving differential equations that are based on conservations laws, physical principles and/or rheological behaviors for a particular situation. However, there remain many complex physical and hydrodynamic processes (as turbulence, interactions between water particles, etc) that have eluded description or explanation. Take for example the lack of reliable and efficient models (admitting fewer variables) for describing turbulence or the fluctuating motion, wave propagation. Despite numerous studies over the past 200 years, the dynamics of coastal flows remain an open problem. We would still like to unify the processes involved to get closer to the reality observed in our oceans, lakes, rivers and so on.

- **Shallow water models with turbulence: high-order shallow water equations**

It is always interesting to learn about the many complex hydrodynamic processes as turbulence observed in coastal zones [12], [13], [14], [15]. The question that arises is how to couple and unify them in a mathematical model? An even more important question is how to model, understand and explain the turbulence processes and its effect in coastal environment problems ? A lack of information about turbulence processes and its consequence, can affect not only the mathematical modeling of the physical problem, but also the way in which the issues at stake in the model are understood. To address turbulence phenomenon in shallow water region we can account the second-order terms associated to the water velocity. This methodology has firstly introduced in [16] for homogeneous flow and in [17] for stratified liquid. This model has been improved by [18] for homogeneous flows (see also [13]) and by [15] for general nonhomogeneous flows with density variation and other mechanisms (see also [19]). The system obtained by [17] named shear shallow water model contains a second order term (or a symmetric tensor admitting three components) that represents the turbulence but is not always hyperbolic and requires others hypotheses associated to this tensor [13]. The implementation of the shear shallow water model is contingent upon the selection of the components of this tensor, a factor that has yet to be of significant interest in practical applications. This model encompasses six unknown variables, is regarded as a binding one, and may become more computationally demanding in a real-world implementation, even when employing precise numerical schemes. An alternative approach to account turbulence involves considering a two-layer shallow water equations, such methodology has recently been introduced in [20] and studied numerically by [21]. The model is unconditionally hyperbolic, yet it necessitates ad hoc assumptions pertaining to the thickness of each layer and the terms coupling both layers. There is no formal derivation of the two-layer shallow water equations; its derivation is also based on ad hoc assumptions.

Then, the existing shallow water models accounting the turbulence (as those developed in [20], [18] [13], [17]), can become sometimes difficult to implement in real applications. It is always desirable to have a more realistic mathematical model having fewer equations, fewer variables. In this article, we develop a new simple long-wave mathematical model of shallow flows that incorporates turbulence (by the second order velocity terms). The 2D model admits a formal derivation without ad hoc assumptions and is unconditionally hyperbolic. It contains five variables associated to five equations.

The proposed high-order model is easy to implement and to study while different from those developed in [17] and extended in [18] (see also [20]) including an additional equation. The model can easily apply to several environment problems (as sediment transport, etc.). In one dimensional, the proposed long-wave high-order model coincides with the high-order models developed by [17] but in two-dimensional case it strongly different.

We demonstrate how fluctuations in velocity in terms of distortion velocity can be employed to account for turbulence in a shallow water system. We then derive a new extension of shallow water equations that account for the turbulence. We show how the turbulence in shallow water context has a dispersive and vortical nature. A gas dynamic analogy is proved in this work. The hyperbolicity of the high order model is demonstrated and some other relevant properties of the introduced model are exposed in details. The objective of this work is to provide a new class of conservation law equations based on second approximation long wave theory. The set of equations describe the motion of a fluid in shallow environment.

The rest of paper is organized as follows. The section 2 exposes a formal derivation of the proposed model. We present also a gas dynamic analogy for the introduced model. The section 3 presents the mathematical properties of the model. In this section, the eigenstructure by a splitting technique is proposed. Additionally, the jump relations and the moving steady states are studied.

## 2. The mathematical model

We consider the time evolution of a water in a (sufficiently smooth) domain $\Omega \subset \mathbb{R}^2$ on a time interval $[0,T]$ with $T > 0$. We write $Q_T = \Omega \times ]0,T]$ and $\Sigma = \Gamma \times ]0,T]$ where $\Gamma = \partial \Omega$

Let $\rho : Q_T \cup \Sigma \to \mathbb{R}$ denote the water density. Let $U : Q_T \cup \Sigma \to \mathbb{R}^3$ denote the three-dimensional water velocity.

### 2.1 Definitions and assumptions

- **Definitions**

We introduce here the vertical average of a function $\psi(\mathbf{x}, z, t)$, $\mathbf{x} = (x, y) \in \mathbb{R}^2, z \in \mathbb{R}, t \in \mathbb{R}_+^*$ by:

$$\overline{\psi}(\mathbf{x},t) = \frac{1}{h} \int_I \psi(\mathbf{x},z,t) dz , \qquad (1)$$

where $I = [Z_b, \eta]$ is the vertical integration.

We define the fluctuation with respect to the average $\psi'(\mathbf{x},z,t)$ given by $\psi' = \psi - \overline{\psi}$ and clearly the average of fluctuation $\psi'$ is zero.

The Leibniz's relations are also used to derive the new density-stratified liquid model:

$$\frac{\partial h\bar{\psi}}{\partial a} = \frac{\partial}{\partial a}\int_I \psi dz = \int_I \frac{\partial \psi}{\partial a}dz - \psi(\eta)\frac{\partial \eta}{\partial a} + \psi(Z_b)\frac{\partial Z_b}{\partial a}, \quad a = x, y \qquad (2)$$

and

$$\frac{\partial h\bar{\psi}}{\partial t} = \frac{\partial}{\partial t}\int_I \psi dz = \int_I \frac{\partial \psi}{\partial t}dz - \psi(\eta)\frac{\partial \eta}{\partial t} + \psi(Z_b)\frac{\partial Z_b}{\partial t}. \qquad (3)$$

- **Assumptions**

The following assumptions are used.

- Long waves propagating assumption $\varepsilon = \frac{H}{L} \ll 1$, where $H$ and $L$ are two scale length characteristics in vertical and horizontal directions respectively.
- The pressure distribution is assumed hydrostatic. There no fluctuation pressure and the mean pressure is of order of $\mathcal{O}(\varepsilon)$.
- The fluid is assumed incompressible and isothermal.
- The velocity fluctuations at the second order are considered.
- The velocity fluctuations correlate and become significant and modify the motion of the fluid while arising an additional motion (named fluctuating motion).
- The fluctuating motion is associated to an additional velocity which transported by the mean velocity.
- The velocity of the fluctuating motion and the velocity of the mean flow have the same order.

## 2.2  Basis equations

We consider the Navier-Stokes equations for nonhomogeneous flow given by:

$$\frac{\partial \rho}{\partial t} + \frac{\partial \rho u_j}{\partial x_j} = 0, \quad \frac{\partial \rho u_i}{\partial t} + \frac{\partial \left(\rho u_i u_j\right)}{\partial x_j} + \frac{\partial p}{\partial x_i} = \mathcal{F}_i, \quad \frac{\partial u_i}{\partial x_i} = 0 \qquad (4)$$

where $x_i = (x, y, z)$ are the 3D coordinates, $u_i$, $i = 1, 2, 3$ or $U = u_i = (u, v, w) = (\mathbf{u}, w)$ are the 3D fluid velocity components, $\rho$ is the water density and $p$ is the pressure term. The term $\mathcal{F}_i$ represents the forces that act in the water flow. The hydrostatic assumption provides an analytical formulation of pressure based on the constant atmospheric pressure and the vertical water column. This assumption involves the neglect of the fluid vertical acceleration in the flow and the z-direction momentum equation becomes:

$$\frac{\partial p}{\partial z} = -\rho g \qquad (5)$$

Considering that the water density is constant, the hydrostatic approximation used in the modeling leads to:

$$\frac{\partial p}{\partial z} = -\rho g \Rightarrow \int_z^\eta \frac{\partial p}{\partial z} d\chi = -\int_z^\eta \rho g d\chi \Rightarrow p(\eta) - p(z) = -\rho g(\eta - z)$$

$$\Rightarrow p(x,y,z) = p(\eta) + \rho g(\eta - z), \quad p(\eta) = p_{atm} = cste$$

The pressure gradient is calculated as:

$$\nabla p = \nabla \rho g(\eta - z) = \rho g \nabla \eta - g(\eta - z) \nabla \rho, \tag{6}$$

and when the density is constant we have $\frac{1}{\rho} \nabla p = g \nabla \eta$

Here, $\nabla$ is gradient operator $\nabla = \left(\frac{\partial}{\partial x}, \frac{\partial}{\partial y}\right)$.

### 2.3 Boundary and kinematic conditions

A position of a fluid particle in a domain $\Omega \subset \mathbb{R}^2$ at the time $t > 0$ is defined as follows:

$$\mathcal{X} = \{(x,y,z), \ (x,y) \in \mathbb{R}^2, \ Z_b(x,y,t) \leq z \leq \eta(x,y,t), \ t \in \mathbb{R}_+^*\},$$

where $\eta$ is a smooth function. A point on the free surface writes: $M_s(x,y,z,t) = z - \eta(x,y,t)$. Similarly, a point at the bed interface writes: $M_b = Z_b(x,y,t) - z$. The normal vectors outward to each interface associated to these curves write respectively $\mathbf{n}_s = \frac{\nabla M_s}{|\nabla M_s|}$ and $\mathbf{n}_b = \frac{\nabla M_b}{|\nabla M_b|}$.

It is assumed that any fluid particle that is on the free surface or on the bottom at the initial instant will remain so at all subsequent instants. The kinematic boundary conditions are therefore given by: $M_s(x,y,z,t) = 0$ and $M_b(x,y,z,t) = 0$, that gives:

$$\frac{DM_{i=s,b}}{Dt} = 0, \quad \text{i.e.} \quad \frac{\partial M_s}{\partial t} + (\mathbf{v}^s . \nabla)(M_s) = 0, \quad \frac{\partial M_b}{\partial t} + (\mathbf{v}^b . \nabla)(M_b) = 0, \tag{7}$$

where $\mathbf{v}^s$, $\mathbf{v}^b$ are the velocity of a point on the free surface respectively on the bottom interface. Here, we assume that the bottom is flat (constant). Particularly, we have $Z_b(x,y,t) = 0$.

### 2.4 Derivation of the HSWE.

To derive the model we integrate the nonhomogeneous Navier-Stokes equations (4) with respect to $z$ from 0 to $h$, In what follows, we will expose some steps of a two-velocity shallow water model. Let consider that $\mathbf{u} = \bar{\mathbf{u}} + (\mathbf{u} - \bar{\mathbf{u}})$. We have the following representation of the integrals:

$$\int_I \mathbf{u} dz = \int_I \bar{\mathbf{u}} dz + \int_I (\mathbf{u} - \bar{\mathbf{u}}) dz = h\bar{\mathbf{u}} + \mathcal{O}(\varepsilon), \tag{8}$$

$$\int_I \mathbf{u} \otimes \mathbf{u}\, dz = \int_I \overline{\mathbf{u}} \otimes \overline{\mathbf{u}}\, dz + \overline{\mathbf{u}} \otimes \int_I (\mathbf{u}-\overline{\mathbf{u}})dz + \int_I (\mathbf{u}-\overline{\mathbf{u}})dz \otimes \overline{\mathbf{u}} + \int_I (\mathbf{u}-\overline{\mathbf{u}}) \otimes (\mathbf{u}-\overline{\mathbf{u}})dz + \mathcal{O}(\varepsilon^2), \quad (9)$$

where, $\mathbf{m} \otimes \mathbf{n}$ denotes the second order tensor with the components $(\mathbf{m} \otimes \mathbf{n})_{ij} = m_i n_j$.

If we consider only term of order $\mathcal{O}(\varepsilon)$ then we can derive form (8) and (9) the classical shallow water equations or first order long wave model:

$$\begin{cases} \dfrac{\partial h}{\partial t} + \nabla \cdot (h\overline{\mathbf{u}}) = 0 \\ \dfrac{\partial h\overline{\mathbf{u}}}{\partial t} + \nabla \cdot (h\overline{\mathbf{u}} \otimes \overline{\mathbf{u}}) + \dfrac{1}{\rho}\overline{\nabla p} = h\overline{\mathbf{F}} + \mathcal{O}(\varepsilon) \end{cases} \quad (10)$$

Here, the fluctuating term $\int_I (\mathbf{u}-\overline{\mathbf{u}}) \otimes (\mathbf{u}-\overline{\mathbf{u}})dz = \mathcal{O}(\varepsilon^{2\beta})$, $0 < \beta < 1$ are neglected since in first approximation long-wave theory, the term of order $\mathcal{O}(\varepsilon^2)$ are smaller than $\mathcal{O}(\varepsilon)$. In second order approximation long wave theory, the term of order $\mathcal{O}(\varepsilon^2)$ can be considered. Then according to the above assumptions the fluctuating term are small for small value of $\varepsilon$ and their value far exceeds the error of the derivation of the long-wave first approximation model (10). This observation is also made in the work of [16]. So when we will consider the term of order such that $\mathcal{O}(\varepsilon^{2\beta}) \gg \mathcal{O}(\varepsilon^2)$, we will naturally account the second order terms $\int_I (\mathbf{u}-\overline{\mathbf{u}}) \otimes (\mathbf{u}-\overline{\mathbf{u}})dz$. According to the fact that the fluctuating motion is strongly associated to a velocity (induced by the turbulence) which have the same order than the mean velocity, we will consider the fluctuations terms $\int_I (\mathbf{u}-\overline{\mathbf{u}}) \otimes (\mathbf{u}-\overline{\mathbf{u}})dz$ as a velocity tensor. This leads to account the turbulence via this new velocity tensor.

In light of these considerations, the classical long wave theory of Saint-Venant is naturally extended, thereby providing a new class of shallow water equations in second order approximation. The second order term accounts for turbulence phenomena.

This phenomenon has been observed in a variety of coastal environments, particularly in the case of a dam break, where a rapid flow over an erodible bottom occurs. In such instances, the turbulence in the main direction of flow exerts a dominant influence on the motion of the water.

$$\int_I (\mathbf{u}-\overline{\mathbf{u}}) \otimes (\mathbf{u}-\overline{\mathbf{u}})dz = h\hat{\mathbf{u}} \otimes \hat{\mathbf{u}}, \quad \hat{\mathbf{u}} = (\hat{u}, \hat{v}), \quad \hat{u}^2 = \overline{(u-\overline{u})(u-\overline{u})}, \quad \hat{v}^2 = \overline{(v-\overline{v})(v-\overline{v})} \quad (11)$$

where $\hat{\mathbf{u}} = \mathcal{O}(\mathbf{u}^*)$, is the horizontal distortion (or fluctuating) water velocity vector along the vertical, given in main flow direction $Ox$ and $Oy$.

The fluctuating motion velocity vector $\hat{\mathbf{u}}$ is of particular interest in the context of coastal engineering applications and dam management, as it allows for the description of a hydrodynamic situation where fluctuations in the shear plane flow direction can be neglected.

Then, (9) becomes:

$$\int_I \mathbf{u} \otimes \mathbf{u}\, dz = h\bar{\mathbf{u}} \otimes \bar{\mathbf{u}} + h\hat{\mathbf{u}} \otimes \hat{\mathbf{u}} + \mathcal{O}(\varepsilon^2). \tag{12}$$

In the same sense, we have:

$$\begin{aligned}\int_I \mathbf{u} \otimes \mathbf{u} \otimes \mathbf{u}\, dz &= \int_I \bar{\mathbf{u}} \otimes \bar{\mathbf{u}} \otimes \bar{\mathbf{u}}\, dz + \int_I \bar{\mathbf{u}} \otimes (\mathbf{u}-\bar{\mathbf{u}}) \otimes (\mathbf{u}-\bar{\mathbf{u}})\, dz + \int_I (\mathbf{u}-\bar{\mathbf{u}}) \otimes (\mathbf{u}-\bar{\mathbf{u}}) \otimes \bar{\mathbf{u}}\, dz + \\ &\quad \int_I (\mathbf{u}-\bar{\mathbf{u}}) \otimes \bar{\mathbf{u}} \otimes (\mathbf{u}-\bar{\mathbf{u}})\, dz + \int_I (\mathbf{u}-\bar{\mathbf{u}}) \otimes (\mathbf{u}-\bar{\mathbf{u}}) \otimes (\mathbf{u}-\bar{\mathbf{u}})\, dz, \\ &= h\bar{\mathbf{u}} \otimes \left((\bar{\mathbf{u}} \otimes \bar{\mathbf{u}}) + (\hat{\mathbf{u}} \otimes \hat{\mathbf{u}})\right) + h(\hat{\mathbf{u}} \otimes \hat{\mathbf{u}}) \otimes \bar{\mathbf{u}} + h\bar{\mathbf{u}} \otimes (\hat{\mathbf{u}} \otimes \hat{\mathbf{u}}) + \\ &\quad \int_I (\mathbf{u}-\bar{\mathbf{u}}) \otimes (\mathbf{u}-\bar{\mathbf{u}}) \otimes (\mathbf{u}-\bar{\mathbf{u}})\, dz + \mathcal{O}(\varepsilon^{2\beta}) + \mathcal{O}(\varepsilon^{3\beta}) \end{aligned} \tag{13}$$

where the third order fluctuation terms $\int_I (\mathbf{u}-\bar{\mathbf{u}}) \otimes (\mathbf{u}-\bar{\mathbf{u}}) \otimes (\mathbf{u}-\bar{\mathbf{u}})\, dz$ are computed using the Fick's law. To account the Reynolds dissipation effect we formulate the third order fluctuations as a diffusion term to produce dissipation of the averaged kinetic energy according to Fick law's. One has:

$$\nabla.\left(\int_I (\mathbf{u}-\bar{\mathbf{u}}) \otimes (\mathbf{u}-\bar{\mathbf{u}}) \otimes (\mathbf{u}-\bar{\mathbf{u}})\, dz\right) \approx \nabla.\left(\overline{\mathbf{u}' \otimes (\mathbf{u}' \otimes \mathbf{u}')}\right) = -\nabla\left(\varepsilon_s \nabla.(\hat{\mathbf{u}} \otimes \hat{\mathbf{u}})\right) = \frac{\partial}{\partial x_k}\left(\varepsilon_s \widehat{u_k} \frac{\partial \hat{u_i}}{\partial x_i}\right),$$

where $\hat{u}_i = (\hat{u}, \hat{v})$ and where the term $\varepsilon_s$ is a coefficient depending on Schmidt number and given in [14].

The tensor $\hat{\mathbf{u}} \otimes \hat{\mathbf{u}}$ has the order $|\hat{\mathbf{u}} \otimes \hat{\mathbf{u}}| = \mathcal{O}(\varepsilon^{2\beta})$ with $|\mathbf{u}-\bar{\mathbf{u}}| = \mathcal{O}(\varepsilon^{\beta})$.

If $\beta < 1$, then $|\hat{\mathbf{u}} \otimes \hat{\mathbf{u}}| > \mathcal{O}(\varepsilon^2)$. According to the second order long wave approximation theory, we can consider the small quantities $\mathcal{O}(\varepsilon^{2\beta}) \gg \mathcal{O}(\varepsilon^2)$. In this case, the term order terms $\int_I (\mathbf{u}-\bar{\mathbf{u}}) \otimes (\mathbf{u}-\bar{\mathbf{u}}) \otimes (\mathbf{u}-\bar{\mathbf{u}})\, dz = \mathcal{O}(\varepsilon^{3\beta})$ can be neglected since $\mathcal{O}(\varepsilon^{2\beta}) \ll \mathcal{O}(\varepsilon^2)$ for $\beta < 1$.

Now, we can derive a simple high-order shallow water model using the above integration relations.

- **Averaged mass conservation equation**

By averaging the free divergence equation and combined with Leibniz relations (2)-(3), and the boundary conditions we obtain:

$$\frac{\partial h}{\partial t} + \nabla.(h\bar{\mathbf{u}}) = 0. \tag{14}$$

- **Averaged momentum conservation equation**

We will design a new momentum equation that includes the effect of turbulence via a new term. By averaging the momentum equation, we obtain:

$$\overline{\frac{\partial \mathbf{u}}{\partial t}} + \overline{\nabla.(\mathbf{u} \otimes \mathbf{u})} + \overline{\frac{\partial (\mathbf{u}u_3)}{\partial z}} + \overline{\frac{\nabla p}{\rho}} = h\overline{\mathbf{F}} \Rightarrow \overline{\frac{\partial \mathbf{u}}{\partial t}} + \overline{\nabla.(\mathbf{u} \otimes \mathbf{u})} + \overline{\frac{\partial (\mathbf{u}u_3)}{\partial z}} + \overline{\frac{\nabla p}{\rho}} = h\overline{\mathbf{F}} \qquad (15)$$

Using the Leibniz relations given by (2)-(3) and the vertical integration (9) we have:

$$\int_I \nabla.(\mathbf{u} \otimes \mathbf{u}) dz = \nabla.\left(h\overline{\mathbf{u}} \otimes \overline{\mathbf{u}}\right) + (\mathbf{u} \otimes \mathbf{u})(\eta) \nabla \eta - (\mathbf{u} \otimes \mathbf{u})(Z_b) \nabla Z_b.$$

kinematic conditions and neglecting the vertical free surface and vertical bottom velocity, we obtain the new averaged momentum conservation law equations written as:

$$\frac{\partial h\overline{\mathbf{u}}}{\partial t} + \nabla.\left(h\overline{\mathbf{u}} \otimes \overline{\mathbf{u}} + h\hat{\mathbf{u}} \otimes \hat{\mathbf{u}}\right) + \frac{1}{\rho} \nabla p = h\overline{\mathbf{F}}, \qquad (16)$$

where the term $\overline{\dfrac{1}{\rho} \nabla p}$ reads $\overline{\dfrac{1}{\rho} \nabla p} = \nabla\left(\dfrac{gh^2}{2}\right)$. We observe that the momentum conservation equations account both mean motion $\nabla.\left(h\overline{\mathbf{u}} \otimes \overline{\mathbf{u}}\right)$ and fluctuating motion $\nabla.\left(h\hat{\mathbf{u}} \otimes \hat{\mathbf{u}}\right)$.

- **Averaged kinetic energy equations**

The kinetic energy equation can be derived from the momentum equations. To provide an averaged kinetic equation, we consider the equation given by:

$$\frac{D\mathbf{u}}{Dt} + \frac{\nabla p}{\rho} = \mathbf{F}, \quad \frac{D}{Dt} = \frac{\partial .}{\partial t} + (\mathbf{u}.\nabla). \qquad (17)$$

By multiplying by $\mathbf{u}$ at left and right of the equation and by summing both obtained equations we obtain:

$$\frac{D\mathbf{u}}{Dt} \otimes \mathbf{u} + \mathbf{u} \otimes \frac{D\mathbf{u}}{Dt} + \frac{\partial (\mathbf{u} \otimes \mathbf{u}u_3)}{\partial z} + \mathbf{u} \otimes \frac{\nabla p}{\rho} + \frac{\nabla p}{\rho} \otimes \mathbf{u} = \mathbf{u} \otimes \mathbf{F} + \mathbf{F} \otimes \mathbf{u},$$

By averaging kinetic equation, we obtain:

$$\overline{\frac{D\mathbf{u}}{Dt} \otimes \mathbf{u}} + \overline{\mathbf{u} \otimes \frac{D\mathbf{u}}{Dt}} + \overline{\mathbf{u} \otimes \frac{\nabla p}{\rho}} + \overline{\frac{\nabla p}{\rho} \otimes \mathbf{u}} + \overline{\mathbf{u} \otimes \frac{\partial (\mathbf{u}u_3)}{\partial z}} + \overline{\frac{\partial (\mathbf{u}u_3)}{\partial z} \otimes \mathbf{u}} = \overline{\mathbf{u} \otimes \mathbf{F}} + \overline{\mathbf{F} \otimes \mathbf{u}},$$
(18)

In following, we assume that $\overline{\mathbf{u} \otimes \mathbf{F}} \approx \overline{\mathbf{u}} \otimes \overline{\mathbf{F}}$ and $\overline{\mathbf{u} \otimes p} \approx \overline{\mathbf{u}} \otimes \overline{p}$.

In this case, one obtain [14]:

$$\overline{\frac{D\mathbf{u}}{Dt} \otimes \mathbf{u}} + \overline{\mathbf{u} \otimes \frac{D\mathbf{u}}{Dt}} + \overline{\mathbf{u}} \otimes \overline{\left(\frac{\nabla p}{\rho}\right)} + \overline{\left(\frac{\nabla p}{\rho}\right)} \otimes \overline{\mathbf{u}} + \overline{\mathbf{u} \otimes \frac{\partial (\mathbf{u}u_3)}{\partial z}} + \overline{\frac{\partial (\mathbf{u}u_3)}{\partial z} \otimes \mathbf{u}} = \overline{\mathbf{u}} \otimes \overline{\mathbf{F}} + \overline{\mathbf{F}} \otimes \overline{\mathbf{u}}, \qquad (19)$$

Considering that:

$$\overline{\frac{D\mathbf{u}}{Dt} \otimes \mathbf{u}} + \overline{\mathbf{u} \otimes \frac{D\mathbf{u}}{Dt}} = \overline{\frac{D(\mathbf{u} \otimes \mathbf{u})}{Dt}} = \overline{\frac{\partial (\mathbf{u} \otimes \mathbf{u})}{\partial t}} + \overline{(\mathbf{u}.\nabla)(\mathbf{u} \otimes \mathbf{u})},$$

where

$$\overline{(\mathbf{u}.\nabla)(\mathbf{u}\otimes\mathbf{u})} = \overline{\sum_{k=1,2} u_k \frac{\partial}{\partial x_k}(\mathbf{u}\otimes\mathbf{u})} = \overline{\sum_{k=1,2} \frac{\partial}{\partial x_k}(\mathbf{u}\otimes\mathbf{u})u_k} = \overline{\nabla(\mathbf{u}\otimes\mathbf{u}).\mathbf{u}}$$
$$= \overline{\nabla.(\mathbf{u}\otimes\mathbf{u}\otimes\mathbf{u})} - \overline{(\mathbf{u}\otimes\mathbf{u})\nabla.\mathbf{u}} = \overline{\nabla.(\mathbf{u}\otimes\mathbf{u}\otimes\mathbf{u})}$$

And taking into account the above Leibniz relations (2)-(3) and also the relation (9), we have respectively:

$$\frac{\partial}{\partial t}\left(h\overline{\mathbf{u}}\otimes\overline{\mathbf{u}} + h\hat{\mathbf{u}}\otimes\hat{\mathbf{u}}\right) = \frac{\partial}{\partial t}\int_I \mathbf{u}\otimes\mathbf{u}\,dz = \int_I \frac{\partial \mathbf{u}\otimes\mathbf{u}}{\partial t}dz - (\mathbf{u}\otimes\mathbf{u})(\eta)\frac{\partial\eta}{\partial t} + (\mathbf{u}\otimes\mathbf{u})(Z_b)\frac{\partial Z_b}{\partial t}.$$

$$\overline{\mathbf{u}\otimes\frac{\partial(uu_3)}{\partial z}} + \overline{\frac{\partial(uu_3)}{\partial z}\otimes\mathbf{u}} = \overline{\frac{\partial(\mathbf{u}\otimes\mathbf{u}u_3)}{\partial z}} \approx (\mathbf{u}\otimes\mathbf{u})(\eta)w(\eta) - (\mathbf{u}\otimes\mathbf{u})(Z_b)w(Z_b)$$

We compute $\overline{\nabla.(\mathbf{u}\otimes\mathbf{u}\otimes\mathbf{u})}$ by using the relations (13) and (2)-(3). Resembling all these relations, we can derive the averaged kinetic energy equation:

$$\frac{\partial(h\widehat{\mathbf{K}})}{\partial t} + \nabla.\left(\frac{h\overline{\mathbf{u}}}{2}\left(\overline{\mathbf{u}}\otimes\overline{\mathbf{u}} + 3\hat{\mathbf{u}}\otimes\hat{\mathbf{u}}\right)\right) + \frac{1}{2}\left(\overline{\frac{\nabla p}{\rho}}\otimes\overline{\mathbf{u}} + \overline{\mathbf{u}}\otimes\overline{\frac{\nabla p}{\rho}}\right) = h\frac{\overline{\mathbf{F}}\otimes\overline{\mathbf{u}}}{2} + h\frac{\overline{\mathbf{u}}\otimes\overline{\mathbf{F}}}{2}, \quad (20)$$

where the kinetic energy $\widehat{\mathbf{K}} = (\widehat{K}_1, \widehat{K}_2)$ in both main flow directions reads $\widehat{\mathbf{K}} = \frac{\hat{\mathbf{u}}\otimes\hat{\mathbf{u}} + \overline{\mathbf{u}}\otimes\overline{\mathbf{u}}}{2}$.

We observe that the averaged kinetic energy accounts both velocities (mean motion velocity $\overline{\mathbf{u}}$ and fluctuating motion velocity $\hat{\mathbf{u}}$). The friction source term is given as [14]:

$$\overline{\mathbf{F}} = -\overline{\mathbf{u}_*^2} = -\overline{\frac{g\varpi^2}{h^{1/3}}\mathbf{u}^2} = -\frac{g\varpi^2}{h^{1/3}}\overline{\mathbf{u}^2} = -\frac{g\varpi^2}{h^{1/3}}\left(\overline{\mathbf{u}}^2 + \hat{\mathbf{u}}^2\right). \quad (21)$$

Then, the friction term depends on water turbulence and this is physically realistic. The turbulence increases the bed shear velocity and then participate to the dissipation.

Then, considering only the terms of order $\mathcal{O}(\varepsilon^{2\beta}) \gg \mathcal{O}(\varepsilon^2)$, the final model reads:

$$\begin{cases} \dfrac{\partial h}{\partial t} + \nabla.\left(h\overline{\mathbf{u}}\right) = 0 \\[2mm] \dfrac{\partial h\overline{\mathbf{u}}}{\partial t} + \nabla.\left(h\overline{\mathbf{u}}\otimes\overline{\mathbf{u}} + h\hat{\mathbf{u}}\otimes\hat{\mathbf{u}}\right) + \dfrac{1}{\rho}\overline{\nabla p} = h\overline{\mathbf{F}} \\[2mm] \dfrac{\partial(h\widehat{\mathbf{K}})}{\partial t} + \nabla.\left(\dfrac{h\overline{\mathbf{u}}}{2}\left(\overline{\mathbf{u}}\otimes\overline{\mathbf{u}} + 3\hat{\mathbf{u}}\otimes\hat{\mathbf{u}}\right)\right) + \dfrac{1}{2}\left(\overline{\dfrac{\nabla p}{\rho}}\otimes\overline{\mathbf{u}} + \overline{\mathbf{u}}\otimes\overline{\dfrac{\nabla p}{\rho}}\right) = h\dfrac{\overline{\mathbf{F}}\otimes\overline{\mathbf{u}}}{2} + h\dfrac{\overline{\mathbf{u}}\otimes\overline{\mathbf{F}}}{2} \end{cases} \quad (22)$$

Fluctuations in velocity are correlated and become greater, and therefore not negligible, in the movement of the fluid. These positive quantities modify the kinematics of the water, accelerating it. We can define a projection to the plane tangent as follows:

$$\left(\hat{\mathbf{u}}\otimes\hat{\mathbf{u}}\right):\left(\overline{\mathbf{u}}\otimes\mathbf{n}\right) = \left(\left(\hat{\mathbf{u}}\otimes\hat{\mathbf{u}}\right)\mathbf{n}\right)_\tau.\overline{\mathbf{u}}_\tau, \quad (23)$$

where $\mathbf{a}_\tau$ denotes the projection of $\mathbf{a}: \partial\Omega \to \mathbb{R}^d$ to the plane tangent to $\partial\Omega$ under consideration. We can introduce also the projection of the normal traction to the tangent plane $\pi = \left(\left(\hat{\mathbf{u}} \otimes \hat{\mathbf{u}}\right)\mathbf{n}\right)_\tau$.

The model has five equation (one for mass conservation equation, two for momentum conservation equations and two averaged kinetic energy equations).

In the model of Tsekunov (that is the first high order shallow water model), the second order terms read $\int_I (\mathbf{u} - \bar{\mathbf{u}}) \otimes (\mathbf{u} - \bar{\mathbf{u}}) dz = P,$ where $P = \left(P_{ij}\right)_{1 \leq i,j \leq 2}$ is a tensor such that:

$$P_{ij} \geq 0 \text{ and } \det\left(P_{ij}\right) = P_{11} P_{12} - P_{12}^2 \geq 0 \tag{24}$$

and leads to the equation following [16]:

$$\begin{cases} \dfrac{\partial h}{\partial t} + \nabla \cdot \left(h \bar{\mathbf{u}}\right) = 0 \\ \dfrac{\partial h \bar{\mathbf{u}}}{\partial t} + \nabla \cdot \left(h \bar{\mathbf{u}} \otimes \bar{\mathbf{u}} + P\right) + \overline{\dfrac{1}{\rho} \nabla p} = 0 \\ \dfrac{\partial \left(h \bar{\mathbf{u}} \otimes \bar{\mathbf{u}} + P\right)}{\partial t} + \nabla \cdot \left(h \bar{\mathbf{u}} \left(\bar{\mathbf{u}} \otimes \bar{\mathbf{u}} + P\right) + P \otimes \bar{\mathbf{u}} + \bar{\mathbf{u}} \otimes P\right) + \left(\overline{\dfrac{\nabla p}{\rho}} \otimes \bar{\mathbf{u}} + \bar{\mathbf{u}} \otimes \overline{\dfrac{\nabla p}{\rho}}\right) = 0 \end{cases} \tag{25}$$

The model developed by Tsekunov (25) has six equations (one for mass conservation equation, two for momentum conservation equations and three averaged kinetic energy equations). In one-dimensional case, the proposed model (22) and the model (25) coincide. The numerical implementation of both models yielded comparable results (see Ngatcha et al., [22]). The model (25) has been extended in several works including those of Tsekunov [17], Richards and Gavrilyuk [18], Ivanova [23], Gavrilyuk et al., [13] and recently in a new general theory of sediment transport developed by Ngatcha and Nkonga [15] (see also Ngatcha et al., [19]). These extensions provide conditionally hyperbolic models under the aforementioned hypothesis (24). Several numerical tests provide interesting results only when the term $P_{12}$ is small or tends to zero.

We can derive equations for the turbulent energy and the fluctuating motion velocity from the equation given by (20). To achieve that, we rewrite as follows:

$$\dfrac{\partial}{\partial t}\left(h\hat{\mathbf{u}} \otimes \hat{\mathbf{u}}\right) + \dfrac{\partial}{\partial t}\left(h\bar{\mathbf{u}} \otimes \bar{\mathbf{u}}\right) + \nabla \cdot \left(h\bar{\mathbf{u}} \otimes \left(\bar{\mathbf{u}} \otimes \bar{\mathbf{u}}\right)\right) + \nabla \cdot \left(3h\bar{\mathbf{u}} \otimes \left(\hat{\mathbf{u}} \otimes \hat{\mathbf{u}}\right)\right) + \left(\overline{\dfrac{\nabla p}{\rho}} \otimes \bar{\mathbf{u}} + \bar{\mathbf{u}} \otimes \overline{\dfrac{\nabla p}{\rho}}\right) = h\bar{\mathbf{F}} \otimes \bar{\mathbf{u}} + h\bar{\mathbf{u}} \otimes \bar{\mathbf{F}} \cdot \tag{26}$$

We use the following relations:

$$\nabla \cdot \left(h\bar{\mathbf{u}} \otimes \left(\bar{\mathbf{u}} \otimes \bar{\mathbf{u}}\right)\right) = \nabla\left(\bar{\mathbf{u}} \otimes \bar{\mathbf{u}}\right) \cdot h\bar{\mathbf{u}} + \left(\bar{\mathbf{u}} \otimes \bar{\mathbf{u}}\right) div\left(h\bar{\mathbf{u}}\right), \tag{27}$$

$$\overline{div\left(\mathbf{u}' \otimes \bar{\mathbf{u}} \otimes \mathbf{u}'\right)} = \dfrac{\partial}{\partial x_\beta}\left(\bar{\mathbf{u}}_\lambda \int_I \mathbf{u}'_\beta \otimes \mathbf{u}'_\beta dz\right) = div\left(h\hat{\mathbf{u}} \otimes \hat{\mathbf{u}}\right) \otimes \bar{\mathbf{u}} + h\hat{\mathbf{u}} \otimes \hat{\mathbf{u}}\left(\nabla \bar{\mathbf{u}}\right)^T, \tag{28}$$

$$\nabla \cdot \left(h\bar{\mathbf{u}} \otimes \left(\hat{\mathbf{u}} \otimes \hat{\mathbf{u}}\right)\right) = \nabla\left(\hat{\mathbf{u}} \otimes \hat{\mathbf{u}}\right) \cdot h\bar{\mathbf{u}} + \left(\hat{\mathbf{u}} \otimes \hat{\mathbf{u}}\right) div\left(h\bar{\mathbf{u}}\right). \tag{29}$$

More generally, we have:

$$[(a \otimes b).\nabla]c + (c.\nabla)(a \otimes b) = \nabla(a \otimes b).c, \quad a,b,c \in \mathbb{R}^n, \qquad (30)$$

$$\nabla(a \otimes b).c = div(a \otimes b \otimes c) - a \otimes b\, div(c), \qquad (31)$$

where $\nabla(a \otimes b).c = \sum_{j=1}^{2} \dfrac{\partial}{\partial x_j}(a \otimes b)c_j$.

We also have:

$$\frac{\partial}{\partial t}\left(h\hat{\mathbf{u}} \otimes \hat{\mathbf{u}}\right) = \hat{\mathbf{u}} \otimes \hat{\mathbf{u}}\frac{\partial h}{\partial t} + h\frac{\partial}{\partial t}\left(\hat{\mathbf{u}} \otimes \hat{\mathbf{u}}\right) = \hat{\mathbf{u}} \otimes \hat{\mathbf{u}}\frac{\partial h}{\partial t} + h\left(\frac{\partial \hat{\mathbf{u}}}{\partial t} \otimes \hat{\mathbf{u}} + \hat{\mathbf{u}} \otimes \frac{\partial \hat{\mathbf{u}}}{\partial t}\right) \qquad (32)$$

We notice that $b.\nabla a$ and $(b.\nabla)a$ have the same role. We have also $a.b^T = b.a$. If $b$ is a non-zero column vector with $b = (b_1, b_2, \ldots, b_n)^T$ then $ab^T = (ab_1, ab_2, \ldots, ab_n)^T$.

We use the fact that:

$$\bar{\mathbf{u}} \otimes \left(\frac{\partial h\bar{\mathbf{u}}}{\partial t} + \nabla.(h\bar{\mathbf{u}} \otimes \bar{\mathbf{u}} + h\hat{\mathbf{u}} \otimes \hat{\mathbf{u}}) + \overline{\frac{1}{\rho}\nabla p} - h\bar{\mathbf{F}}\right) + \left(\frac{\partial h\bar{\mathbf{u}}}{\partial t} + \nabla.(h\bar{\mathbf{u}} \otimes \bar{\mathbf{u}} + h\hat{\mathbf{u}} \otimes \hat{\mathbf{u}}) + \overline{\frac{1}{\rho}\nabla p} - h\bar{\mathbf{F}}\right) \otimes \bar{\mathbf{u}} = 0 \quad (33)$$

Using the above relations, the kinetic energy relation given by (26) becomes:

$$\frac{\partial}{\partial t}\left(\hat{\mathbf{u}} \otimes \hat{\mathbf{u}}\right) + \nabla.\left(\hat{\mathbf{u}} \otimes \hat{\mathbf{u}}\right)\bar{\mathbf{u}} + \hat{\mathbf{u}} \otimes \hat{\mathbf{u}}\left(\nabla \bar{\mathbf{u}}\right)^T + \nabla \bar{\mathbf{u}}\left(\hat{\mathbf{u}} \otimes \hat{\mathbf{u}}\right) = 0.$$

That leads to:

$$\hat{\mathbf{u}} \otimes \left(\frac{d\hat{\mathbf{u}}}{dt} + (\nabla \bar{\mathbf{u}})\hat{\mathbf{u}}\right) + \left(\frac{d\hat{\mathbf{u}}}{dt} + (\nabla \bar{\mathbf{u}})\hat{\mathbf{u}}\right) \otimes \hat{\mathbf{u}} = 0. \qquad (34)$$

We obtain the equation the velocity $\hat{\mathbf{u}}$ as follows:

$$\frac{d\hat{\mathbf{u}}}{dt} + (\hat{\mathbf{u}}.\nabla)\bar{\mathbf{u}} = 0.$$

That is equivalent to:

$$\begin{cases} \partial_t \hat{u} + \partial_x(\hat{u}\bar{u}) + \bar{v}\partial_y \hat{u} + \hat{v}\partial_y \bar{u} = 0 \\ \partial_t \hat{v} + \bar{u}\partial_x \hat{v} + \hat{u}\partial_x \bar{v} + \partial_y(\bar{v}\hat{v}) = 0 \end{cases} \qquad (35)$$

The final long-wave second-order shallow water model with a free boundary in $\Omega \times ]0,T]$ reads:

$$\frac{\partial h}{\partial t}+\frac{\partial h\overline{u}}{\partial x}+\frac{\partial h\overline{v}}{\partial y}=0,$$

$$\frac{\partial h\overline{u}}{\partial t}+\frac{\partial}{\partial x}\left(h\hat{u}^2+h\overline{u}^2\right)+\frac{\partial}{\partial y}\left(h\hat{u}\hat{v}+h\overline{u}\overline{v}\right)+gh\frac{\partial h}{\partial x}=h\overline{F}_1,$$

$$\frac{\partial h\overline{v}}{\partial t}+\frac{\partial}{\partial y}\left(h\hat{u}\hat{v}+h\overline{u}\overline{v}\right)+\frac{\partial}{\partial y}\left(h\hat{v}^2+h\overline{v}^2\right)+gh\frac{\partial h}{\partial y}=h\overline{F}_2, \quad (36)$$

$$\frac{\partial h\widehat{K}_1}{\partial t}+\frac{\partial}{\partial x}\left(\frac{h\overline{u}}{2}\left(3\hat{u}^2+\overline{u}^2\right)\right)+\frac{\partial}{\partial y}\left(\frac{h\overline{v}}{2}\left(\overline{u}\overline{v}+3\hat{u}\hat{v}\right)\right)+gh\overline{u}\frac{\partial h}{\partial x}=h\overline{u}\overline{F}_1+\frac{\partial}{\partial x}\left(\varepsilon_s\hat{u}\frac{\partial\hat{u}}{\partial x}\right),$$

$$\frac{\partial h\widehat{K}_2}{\partial t}+\frac{\partial}{\partial x}\left(\frac{h\overline{u}}{2}\left(\overline{u}\overline{v}+3\hat{u}\hat{v}\right)\right)+\frac{\partial}{\partial y}\left(\frac{h\overline{v}}{2}\left(3\hat{v}^2+\overline{v}^2\right)\right)+gh\overline{v}\frac{\partial h}{\partial y}=h\overline{v}\overline{F}_2+\frac{\partial}{\partial y}\left(\varepsilon_s\hat{v}\frac{\partial\hat{v}}{\partial y}\right).$$

We joint to this second-order model (36) the initial and boundary conditions defined respectively by:

$$h(\mathbf{x},0)=h^0(\mathbf{x}), \quad u(\mathbf{x},0)=u^0(\mathbf{x}), v(\mathbf{x},0)=v^0(x,y), \quad \widehat{K}_1(\mathbf{x},0)=\widehat{K}_1^{\,0}(\mathbf{x}), \widehat{K}_2(\mathbf{x},0)=\widehat{K}_2^{\,0}(\mathbf{x}) \quad (37)$$

and

$$h\mathbf{n}=0, \quad (h\overline{u})\mathbf{n}=0, \quad (h\overline{v})\mathbf{n}=0, \quad \widehat{K}_1\mathbf{n}=0, \widehat{K}_2\mathbf{n}=0, \text{ on } \partial\Omega, \forall t\in]0,T], \quad (38)$$

where $\mathbf{n}:\partial\Omega\to\mathbb{R}^2$ denotes the mapping that assigns the outward unit normal vector to any $\mathbf{x}\in\partial\Omega$.

In above equation, the term $\dfrac{\partial}{\partial x_k}\left(\varepsilon_s\widehat{u}_k\dfrac{\partial\widehat{u}_i}{\partial x_i}\right)$ represents the turbulence transport. It describes the dissipation phenomena in the sediment transport [14]. The model (36) is given in terms of kinetic energy in main flow directions. By removing the term $\dfrac{\partial}{\partial x_k}\left(\varepsilon_s\widehat{u}_k\dfrac{\partial\widehat{u}_i}{\partial x_i}\right)$ in the modeling, it also possible to write the model in terms of fluctuating motion velocity (or distortion velocities $\hat{u}$, $\hat{v}$ respectively) as follows:

$$\frac{\partial h}{\partial t}+\frac{\partial h\overline{u}}{\partial x}+\frac{\partial h\overline{v}}{\partial y}=0,$$

$$\frac{\partial h\overline{u}}{\partial t}+\frac{\partial}{\partial x}\left(h\hat{u}^2+h\overline{u}^2+\frac{1}{2}gh^2\right)+\frac{\partial}{\partial y}\left(h\hat{u}\hat{v}+h\overline{u}\overline{v}\right)=h\overline{F}_1,$$

$$\frac{\partial h\overline{v}}{\partial t}+\frac{\partial}{\partial y}\left(h\hat{u}\hat{v}+h\overline{u}\overline{v}\right)+\frac{\partial}{\partial y}\left(h\hat{v}^2+h\overline{v}^2+\frac{1}{2}gh^2\right)=h\overline{F}_2, \quad (39)$$

$$\partial_t\hat{u}+\partial_x\left(\overline{u}\hat{u}\right)+\overline{v}\partial_y\hat{u}+\hat{v}\partial_y\overline{u}=0,$$

$$\partial_t\hat{v}+\overline{u}\partial_x\hat{v}+\hat{u}\partial_x\overline{v}+\partial_y\left(\overline{v}\hat{v}\right)=0.$$

Or

$$\frac{\partial h}{\partial t} + \frac{\partial hu}{\partial x} + \frac{\partial hv}{\partial y} = 0,$$

$$\frac{\partial hu}{\partial t} + \frac{\partial}{\partial x}\left(h\hat{u}^2 + h\overline{u}^2 + \frac{1}{2}gh^2\right) + \frac{\partial}{\partial y}\left(h\hat{u}\hat{v} + h\overline{u}\overline{v}\right) = h\overline{\mathbf{F}}_1,$$

$$\frac{\partial hv}{\partial t} + \frac{\partial}{\partial y}\left(h\hat{u}\hat{v} + h\overline{u}\overline{v}\right) + \frac{\partial}{\partial y}\left(h\hat{v}^2 + h\overline{v}^2 + \frac{1}{2}gh^2\right) = h\overline{\mathbf{F}}_2, \qquad (40)$$

$$\partial_t \hat{u} + \partial_x(\overline{u}\hat{u}) + \partial_y(\overline{v}\hat{u}) = \hat{u}\partial_y \overline{v} - \hat{v}\partial_y \overline{u},$$

$$\partial_t \hat{v} + \partial_x(\overline{u}\hat{v}) + \partial_y(\overline{v}\hat{v}) = \hat{v}\partial_x \overline{u} - \hat{u}\partial_x \overline{v}.$$

We joint to this second-order nonconservative model (39) or its conservative form (40) the initial and boundary conditions defined respectively by:

$$h(\mathbf{x},0) = h^0(\mathbf{x}), \quad \overline{u}(\mathbf{x},0) = \overline{u}^0(\mathbf{x}), \quad \overline{v}(\mathbf{x},0) = \overline{v}^0(\mathbf{x}), \quad \hat{u}(\mathbf{x},0) = \hat{u}^0(\mathbf{x}), \quad \hat{v}(\mathbf{x},0) = \hat{v}^0(\mathbf{x}) \qquad (41)$$

and

$$h\mathbf{n} = 0, \quad (h\overline{u})\mathbf{n} = 0, \quad (h\overline{v})\mathbf{n} = 0, \quad \hat{u}\mathbf{n} = 0, \quad \hat{v}\mathbf{n} = 0. \qquad (42)$$

The high order model exposed in Eq. (39) is based on one layer fluid and is strongly different to the familiar one exposed in [16]. It is also different to the model described in [20] and based on two layers fluid. Some differences of the three modellings are exposed below. We will see that the introduced theory is more efficient than both others.

**Remark:** *Difference with existing shallow water models accounting the turbulence*

There are notable distinctions between our modeling approach and those employed in [20] and [16] (see also [13]). The discrepancies can be attributed to differences in assumptions, variables, and derivation methodologies. Our derivation is distinct from that employed in [16] and even in [20]. The model developed by [20] is a bilayer-based model including ad hoc assumptions, while that developed by [16] encompasses a multitude of complexities associated with the variable describing turbulence. Additionally, it can become conditionally hyperbolic.

In the work developed by [17], the additional equation for the evolution of the kinetic energy in the shear plane may prove to be a less useful tool in certain environmental situations. Furthermore, it introduces complications to the physical study of waves and makes it challenging to demonstrate the existence of a weak global solution. With the proposed model, there is no kinetic energy equation in shear plane and this is original and useful in real applications. We can prove the existence of a global weak solution and this is one main originality of this model. The model differs from that developed by [20], which is a bilayer model including several unknown variables. The model developed by [20] assumes that all layers have the same thickness, which may not be the case in real applications. Furthermore, its modeling requires closure relations. In contrast, the proposed one layer genuinely second order model, with its formal derivation, is simple, easy to implement and has fewer variables, which is advantageous for real applications that may involve other complexities, such as space-time variabilities, domain, meshes and method used. The classical shallow water based models based on homogeneous flow content three equations (1 mass conservation equation, 2 momentum conservation equations). The proposed model has five equations (the three classical equations plus 2 additional equations describing the evolution of kinetic energy in the main flow directions). Several possible extensions of the model to other physics remain open. In one-dimensional case, the model (36) in its nonconservative form reads:

$$\begin{cases} \dfrac{\partial h}{\partial t} + \dfrac{\partial h\overline{u}}{\partial x} = 0, \\ \dfrac{\partial h\overline{u}}{\partial t} + \dfrac{\partial}{\partial x}\left(h\hat{u}^2 + h\overline{u}^2\right) + gh\dfrac{\partial h}{\partial x} = h\overline{\mathbf{F}}_1, \\ \dfrac{\partial h\widehat{K}_1}{\partial t} + \dfrac{\partial}{\partial x}\left(\dfrac{h\overline{u}}{2}\left(3\hat{u}^2 + \overline{u}^2\right)\right) + gh\overline{u}\dfrac{\partial h}{\partial x} = h\overline{u}\overline{\mathbf{F}}_1 + \dfrac{\partial}{\partial x}\left(\varepsilon_s \hat{u}\dfrac{\partial \hat{u}}{\partial x}\right). \end{cases} \quad (43)$$

Or using the form (39) we read:

$$\begin{cases} \dfrac{\partial h}{\partial t} + \dfrac{\partial h\overline{u}}{\partial x} = 0, \\ \dfrac{\partial h\overline{u}}{\partial t} + \dfrac{\partial}{\partial x}\left(h\hat{u}^2 + h\overline{u}^2 + \dfrac{1}{2}gh^2\right) = h\overline{\mathbf{F}}_1, \\ \dfrac{\partial \hat{u}}{\partial t} + \dfrac{\partial}{\partial x}\left(\hat{u}\overline{u}\right) = 0. \end{cases} \quad (44)$$

When we remove the turbulence modeling, we retrieve the well-known shallow water equations in nonconservative form given by:

$$\begin{cases} \dfrac{\partial h}{\partial t} + \dfrac{\partial h\overline{u}}{\partial x} = 0, \\ \dfrac{\partial h\overline{u}}{\partial t} + \dfrac{\partial}{\partial x}\left(h\overline{u}^2\right) + gh\dfrac{\partial h}{\partial x} = h\overline{\mathbf{F}}_1, \end{cases} \quad (45)$$

The model given by (45) can describe steady and unsteady (or rapid unsteady) flows in coastal regions but cannot represents accelerated flows that arise during dam breaks as the model given by (43) or by (39) ( or (36)). Particularly, it does not consider the effect of the vortical coherent structures associated to the rapid flow that modify the wave structure of the flow. In addition, any unsteady flow is not necessarily turbulent.

## 2.5. Gas dynamic analogy

From the last equation of (44) we have:

$$\hat{u}\dfrac{\partial \hat{u}}{\partial t} + \hat{u}\dfrac{\partial}{\partial x}\left(\hat{u}\overline{u}\right) = 0 \Rightarrow \dfrac{\partial}{\partial t}\dfrac{\hat{u}^2}{2} + \dfrac{\partial}{\partial x}\left(\overline{u}\dfrac{\hat{u}^2}{2}\right) = 0 \Rightarrow \dfrac{\partial}{\partial t}h\dfrac{\hat{u}^2}{2} + \dfrac{\partial}{\partial x}\left(h\overline{u}\dfrac{\hat{u}^2}{2}\right) = 0$$

$$\Rightarrow \dfrac{\partial h\omega}{\partial t} + \dfrac{\partial}{\partial x}\left(h\overline{u}\omega\right) = 0 \quad (46)$$

where $\omega = \dfrac{\hat{u}^2}{2}$ is the averaged vorticity and the equation (46) represents the production of vorticity $\omega$. In order to estimate the production of vorticity, one could derive an evolution equation for the enstrophy density $\Phi = \dfrac{1}{2}\omega^2$ (vorticity squared) by the dot product of (46) with $\omega$.

$$\Rightarrow \frac{\partial}{\partial t}\left(\frac{\omega^2}{2}\right) + \frac{\partial}{\partial x}\left(\bar{u}\left(\frac{\omega^2}{2}\right)\right) = 0 \Rightarrow \frac{\partial(h\Phi)}{\partial t} + \frac{\partial}{\partial x}(h\bar{u}\Phi) = 0 \Rightarrow \frac{d\Phi}{dt} = 0 \qquad (47)$$

From we observe that $\Phi$ is conserved along of trajectories. The vortex dominates in the compute of the rate enstrophy. The contribution of enstrophy production and destruction rates by vortex depends on the wavenumber. The enstrophy represents a scalar quantity that intrinsically reflects the strength of the vorticity field without a vector implication. The enstrophy can help to measure vortical energy of turbulent structures. The equation (47) can be rewritten as follows:

$$\frac{\partial(h^3\Phi)}{\partial t} + \frac{\partial}{\partial x}(h^3\bar{u}\Phi) = 0 \qquad (48)$$

and the mechanical energy equation becomes

$$\frac{\partial}{\partial t}\left(h\left(\frac{\bar{u}}{2} + \underbrace{h^2\Phi + \frac{gh}{2}}_{e}\right)\right) + \frac{\partial}{\partial x}\bar{u}\left(h\frac{\bar{u}}{2} + h\underbrace{\left(h^2\Phi + \frac{gh}{2}\right)}_{e} + \underbrace{h^3\frac{\Phi}{2} + \frac{gh^2}{2}}_{\tilde{p}}\right) = 0 \quad, \qquad (49)$$

For short we write:

$$\frac{\partial}{\partial t}\left(h\frac{\bar{u}}{2} + he\right) + \frac{\partial}{\partial x}\bar{u}\left(h\frac{\bar{u}}{2} + he + \tilde{p}\right) = 0 \qquad (50)$$

where $e(h,\Phi) = h^2\Phi + \dfrac{gh}{2}$ and $\tilde{p}(h,\Phi) = h^3\dfrac{\Phi}{2} + \dfrac{gh^2}{2}$.

Introducing the specific entropy $s$ such that

$$h^2 T ds = h^2 de + \tilde{p} dh \qquad (51)$$

where $T$ is the temperature

$$T ds = de + \tilde{p} d\left(\frac{1}{h}\right) = h^2 d\omega \qquad (52)$$

If the averaged vorticity is constant then $d\omega = 0 \Rightarrow ds = 0 \Rightarrow s = cste$.

The equation (52) leads to $T = h^2 \dfrac{d\omega}{ds}$ and by letting $s = \ln(\omega^2)$ we find $T = h^2 \dfrac{\omega^2}{2}$.

With these above relations, we have proved a gas dynamic analogy (equations of nonisentropic gas dynamic) for the proposed model given by (43).

The resulting nonisentropic gas dynamic equations read:

$$\begin{cases} \dfrac{\partial \rho}{\partial t} + \dfrac{\partial}{\partial x}\left(\rho \bar{u}\right) = 0, \\ \dfrac{\partial \rho \bar{u}}{\partial t} + \dfrac{\partial}{\partial x}\left(\rho \bar{u}^2 + \tilde{p}\right) = 0, \\ \dfrac{\partial}{\partial t}\left(\rho \dfrac{\bar{u}^2}{2} + \rho e\right) + \dfrac{\partial}{\partial x}\bar{u}\left(\rho \dfrac{\bar{u}^2}{2} + \rho e + \tilde{p}/\rho\right) = 0, \\ e(s,\rho) = \rho^2 \dfrac{e^s}{2} + \dfrac{g\rho}{2}, \quad \tilde{p}(s,\rho) = \rho^3 \dfrac{e^s}{2} + \dfrac{g\rho^2}{2} \end{cases} \quad (53)$$

The two last equations represent the equations of state of gas. The case where the entropy is constant corresponds to isentropic gas dynamic and when $e^s \to 0$, $e(\rho,s) \to \dfrac{g\rho}{2}$, $\tilde{p} \to \dfrac{g\rho^2}{2}$.

Then the internal energy becomes the potential energy and then the system (53) converges to the shallow water system (45).

## 3. Mathematical properties of the high order shallow water model

### 3.1. Total energy and entropy relations

**Total energy equation**

We can find an entropy pair for the model. We show first the following result:

***Proposition: 3.1***

*Consider the third generation sediment transport system given by* (36) *. Any smooth solution of* (36) *satisfy the total energy equation given by:*

$$\dfrac{\partial \mathcal{E}(\mathbf{W})}{\partial t} + \nabla . G(\mathbf{W}) = \overline{\mathbf{F}\mathbf{u}},$$

*where* $\mathcal{E}(\mathbf{W})$ *and* $G(\mathbf{W})$ *are two functions given respectively the mechanical energy and the flux entropy flux given by:*

$$\mathcal{E}(\mathbf{W}) = h\,tr(\widehat{\mathbf{K}}) + \dfrac{gh^2}{2} \quad \text{and} \quad G(\mathbf{W}) = \left(\mathcal{E} + h\left(\hat{\mathbf{u}} \otimes \hat{\mathbf{u}}\right) + \dfrac{gh^2}{2}\right)\bar{\mathbf{u}}, \quad (54)$$

*where* $tr(\widehat{\mathbf{K}}) = \widehat{K}_1 + \widehat{K}_2$.

***Proof***

We consider the water-sediment mass conservation equation and the averaged kinetic energy equation. We multiply the mass conservation by $gh$, we sum the two components of the averaged kinetic energy equations.

**Entropy and energy inequalities**

*Proposition 3.2*

Consider the system given by (36). Any smooth solution of (36) satisfies the equation following:

$$\frac{\partial \mathcal{E}(\mathbf{W})}{\partial t} + \nabla . G(\mathbf{W}) = \overline{\mathbf{F}\mathbf{u}} \tag{55}$$

where $\mathcal{E}(\mathbf{W})$ and $G(\mathbf{W})$ are two functions given respectively by :

$$\mathcal{E}(\mathbf{W}) = h\left(\widehat{K}_1 + \widehat{K}_2\right) + \frac{gh^2}{2}, G(\mathbf{W}) = \left(\mathcal{E} + h\hat{\mathbf{u}} \otimes \hat{\mathbf{u}} + \frac{gh^2}{2}\right)\overline{\mathbf{u}}.$$

*Corollary 3.1*

Consider the system given by (36) and suppose that the following assumption are satisfied

$$\overline{\mathbf{F}\mathbf{u}} \leq 0. \tag{56}$$

Then, any smooth solution of (36) satisfies the entropy inequality

$$\frac{\partial \mathcal{E}(\mathbf{W})}{\partial t} + \nabla . G(\mathbf{W}) \leq 0, \tag{57}$$

where $\mathcal{E}(\mathbf{W})$ and $G(\mathbf{W})$ are two functions given above.

The model is nonconservative but its admits a conservative one given by:

$$\begin{cases} \frac{\partial h}{\partial t} + \nabla .\left(h\overline{\mathbf{u}}\right) = 0 \\ \frac{\partial h\overline{\mathbf{u}}}{\partial t} + \nabla .\left(h\overline{\mathbf{u}} \otimes \overline{\mathbf{u}} + h\hat{\mathbf{u}} \otimes \hat{\mathbf{u}}\right) + \frac{1}{\rho}\nabla p = h\overline{\mathbf{F}} \\ \frac{\partial \mathcal{E}}{\partial t} + \nabla .\left(\mathcal{E} + h\hat{\mathbf{u}} \otimes \hat{\mathbf{u}} + \frac{gh^2}{2}\right)\overline{\mathbf{u}} = h\frac{\overline{\mathbf{F}} \otimes \overline{\mathbf{u}}}{2} + h\frac{\overline{\mathbf{u}} \otimes \overline{\mathbf{F}}}{2} \end{cases} \tag{58}$$

With such a form, we can avoid the difficulties associated with processing non-conservative terms.

### 3.2 Reformulation and hyperbolicity

We can reformulate the system (36) in the form:

$$\frac{\partial \mathbf{W}}{\partial t} + \mathcal{A}_1(\mathbf{W})\frac{\partial \mathbf{W}}{\partial x} + \mathcal{A}_2(\mathbf{W})\frac{\partial \mathbf{W}}{\partial y} = \mathbf{S}(\mathbf{W}), \tag{59}$$

where $\mathcal{A}_1(\mathbf{W}) = \frac{\partial F_1}{\partial x} + \mathbf{B}_1(\mathbf{W})$, $\mathcal{A}_2(\mathbf{W}) = \frac{\partial F_2}{\partial y} + \mathbf{B}_2(\mathbf{W})$.

where $\mathbf{x} = (x,y) \in \Omega \subset \mathbb{R}^2$, $t \in ]0,T]$, $T > 0$. The vector unknowns is a function defined in the following space $\mathbf{W}: \mathbb{R}^2 \times \mathbb{R}^+ \to \mathbb{R}^5$ and the nonconservative vector $\mathbf{B}_{x,y}$ are given by:

$$\mathbf{W} = \left(h, h\bar{u}, h\bar{v}, h\widehat{K}_1, h\widehat{K}_2\right), \mathbf{B}_x = \left(0, gh, 0, gh\bar{u}, 0\right), \mathbf{B}_y = \left(0, 0, gh, 0, gh\bar{v}\right), \tag{60}$$

while the physical flux functions $F_1, F_2 : \mathbb{R}^5 \to \mathbb{R}^5$ are given by:

$$F_1 = \left(h\bar{u}, h\hat{u}^2 + h\bar{u}^2, h\hat{u}\hat{v} + h\bar{u}\bar{v}, \frac{h\bar{u}}{2}\left(\bar{u}^2 + 3\hat{u}^2\right), \frac{h\bar{u}}{2}\left(\bar{u}\bar{v} + 3\hat{u}\hat{v}\right)\right)^t$$

$$F_2 = \left(h\bar{v}, h\hat{u}\hat{v} + h\bar{u}\bar{v}, h\hat{v}^2 + h\bar{v}^2, \frac{h\bar{v}}{2}\left(\bar{u}\bar{v} + 3\hat{u}\hat{v}\right), \frac{h\bar{v}}{2}\left(\bar{v}^2 + 3\hat{v}^2\right)\right)^t \tag{61}$$

or in the form

$$\frac{\partial \mathbf{W}}{\partial t} + \mathcal{A}_\mathbf{n}(\mathbf{W})\nabla \mathbf{W} = \mathbf{S}(\mathbf{W}), \tag{62}$$

where

$\mathcal{A}_\mathbf{n} = \mathcal{A}(\mathbf{W},\mathbf{n}) = \mathcal{A}_1.n_1 + \mathcal{A}_2.n_2 = \nabla F.\mathbf{n} + \mathbf{B}.\mathbf{n}$, where $\mathbf{n} = (n_1, n_2)$ is a normal vector and $\mathbf{B}.\mathbf{n}$ being $\mathbf{B}.\mathbf{n} = \mathbf{B}_1.n_1 + \mathbf{B}_2.n_2 = (0, n_1, n_2, un_1, vn_2)$.

The function $F.\mathbf{n}$ being the flux conservative contributions.

**Hyperbolicity study by a splitting technique**

We consider the 1D nonconservative system given by

$$\begin{aligned}
&\frac{\partial h}{\partial t} + \frac{\partial hu}{\partial x} = 0, \\
&\frac{\partial hu}{\partial t} + \frac{\partial}{\partial x}\left(h\hat{u}^2 + h\bar{u}^2\right) + gh\frac{\partial h}{\partial x} = h\bar{F}_1, \\
&\frac{\partial hv}{\partial t} + \frac{\partial}{\partial x}\left(h\hat{u}\hat{v} + h\bar{u}\bar{v}\right) = h\bar{F}_2, \\
&\frac{\partial h\widehat{K}_1}{\partial t} + \frac{\partial}{\partial x}\left(\frac{h\bar{u}}{2}\left(3\hat{u}^2 + \bar{u}^2\right)\right) + gh\bar{u}\frac{\partial h}{\partial x} = h\bar{u}\bar{F}_1 + \frac{\partial}{\partial x}\left(\varepsilon_s\hat{u}\frac{\partial\hat{u}}{\partial x}\right), \\
&\frac{\partial h\widehat{K}_2}{\partial t} + \frac{\partial}{\partial x}\left(\frac{h\bar{v}}{2}\left(\bar{u}\bar{v} + 3\hat{u}\hat{v}\right)\right) = h\bar{v}\bar{F}_2.
\end{aligned} \tag{63}$$

The system admits its following conservative form

$$\begin{cases} \dfrac{\partial h}{\partial t} + \dfrac{\partial hu}{\partial x} = 0, \\ \dfrac{\partial hu}{\partial t} + \dfrac{\partial}{\partial x}\left(\hat{p} + h\overline{u}^2\right) = h\overline{F}_1, \\ \dfrac{\partial hv}{\partial t} + \dfrac{\partial}{\partial x}\left(h\hat{u}\hat{v} + h\overline{u}\overline{v}\right) = h\overline{F}_2, \\ \dfrac{\partial hE_1}{\partial t} + \dfrac{\partial}{\partial x}\left(h\overline{u}(E_1 + \hat{p})\right) = h\overline{u}\overline{F}_1 + \dfrac{\partial}{\partial x}\left(\varepsilon_s \hat{u}\dfrac{\partial \hat{u}}{\partial x}\right), \\ \dfrac{\partial h\widehat{K}_2}{\partial t} + \dfrac{\partial}{\partial x}\left(\dfrac{h\overline{v}}{2}\left(\overline{uv} + 3\hat{u}\hat{v}\right)\right) = h\overline{v}\overline{F}_2, \end{cases} \quad (64)$$

where we have denoted $\hat{p} = \dfrac{gh^2}{2} + h\hat{u}^2$.

One solve the five equations of the system (64) by several finite volume schemes available in the literature and we compute $\widehat{K}_2$ by the mechanical energy $E_1 = \dfrac{gh}{2} + \widehat{K}_1$. Indeed, we have $\widehat{K}_1 = E_1 - \dfrac{gh}{2}$. The quasi-1D system is hyperbolic and admits three types of wave: a contact discontinuity waves with the velocity $\overline{u}$, accelerated surface gravity waves with the velocity $\overline{u} \pm \sqrt{gh + 3\hat{u}^2}$ and the fluctuating (or distortion) waves of velocity $\overline{u} \pm \hat{u}$. To study the eigenstructure (eigenvalues and eigenvectors) of the system (64), we split in two categories waves subsystems: the accelerated-wave subsystem and the distortion-wave subsystem. For both subsystems we will remove the source terms.

**Accelerated-wave subsystem**

The accelerated-wave subsystem reads:

$$\begin{cases} \dfrac{\partial h}{\partial t} + \dfrac{\partial hu}{\partial x} = 0, \\ \dfrac{\partial hu}{\partial t} + \dfrac{\partial}{\partial x}\left(\hat{p} + h\overline{u}^2\right) = 0, \\ \dfrac{\partial hv}{\partial t} + \dfrac{\partial}{\partial x}\left(h\overline{uv}\right) = 0, \\ \dfrac{\partial \hat{u}}{\partial t} + \dfrac{\partial}{\partial x}\left(\overline{u}\hat{u}\right) = 0, \\ \dfrac{\partial \hat{v}}{\partial t} + \overline{u}\dfrac{\partial \hat{v}}{\partial x} = 0. \end{cases} \quad (65)$$

The system (65) can rewrite as:

$$\frac{\partial \mathbf{W}}{\partial t} + \mathcal{A}(\mathbf{W})\partial_x \mathbf{W} = 0,$$

where $\mathbf{W} = (h, \bar{u}, \hat{u}, \bar{v}, \hat{v})$ and where $\mathcal{A}(\mathbf{W}) = \begin{pmatrix} \bar{u} & h & 0 & 0 & 0 \\ g + \frac{\hat{u}^2}{h} & \bar{u} & 2\hat{u} & 0 & 0 \\ 0 & \hat{u} & \bar{u} & 0 & 0 \\ 0 & 0 & \bar{u} & \bar{u} & 0 \\ 0 & 0 & 0 & \hat{u} & \bar{u} \end{pmatrix}$

The eigenvalues of $\mathcal{A}$ are:

$$\lambda_{1,2,3} = \bar{u}; \lambda_{4,5} = \bar{u} \pm a, \quad a = \sqrt{gh + 3\hat{u}^2} \tag{66}$$

where $\lambda_5$ is the accelerated wave velocity. The eigenvectors associated to these eigenvalues are:

$$R_1 = \begin{pmatrix} h \\ 0 \\ \frac{(gh+\hat{u}^2)}{2\hat{u}} \\ 0 \\ 0 \end{pmatrix}, R_2 = \begin{pmatrix} 0 \\ 0 \\ 0 \\ 1 \\ 0 \end{pmatrix}, R_3 = \begin{pmatrix} 0 \\ 0 \\ 0 \\ 0 \\ 1 \end{pmatrix}, R_4 = \begin{pmatrix} h \\ -a \\ \hat{u} \\ -\bar{u}\hat{u} \\ a \\ 0 \end{pmatrix}, R_5 = \begin{pmatrix} h \\ a \\ \hat{u} \\ \bar{u}\hat{u} \\ a \\ 0 \end{pmatrix}$$

We have:

$$\nabla_\mathbf{W}\lambda_1 \cdot R_1 = 0, \ \nabla_\mathbf{W}\lambda_2 \cdot R_2 = 0, \ \nabla_\mathbf{W}\lambda_3 \cdot R_3 = 0, \ \nabla_\mathbf{W}\lambda_4 \cdot R_4 = \frac{-a^2}{2a} + 3\hat{u}^2, \nabla_\mathbf{W}\lambda_5 \cdot R_5 = \frac{a^2}{2a} + 3\hat{u}^2$$

The eigenvector are linearly independent since $det(R_1, R_2, R_3, R_4, R_5) = -\frac{a^3 h}{\hat{u}} \neq 0, \ \hat{u} > 0$. Then the system is hyperbolic.

**Subsystem for distortion-waves.**

The subsystem for distortion-waves reads:

$$\begin{cases} \dfrac{\partial h}{\partial t} = 0, \\[4pt] \dfrac{\partial hu}{\partial t} = 0, \\[4pt] \dfrac{\partial v}{\partial t} + \dfrac{1}{h}\dfrac{\partial}{\partial x}\left(h\hat{u}\hat{v}\right) = 0, \\[4pt] \dfrac{\partial \hat{u}}{\partial t} = 0, \\[4pt] \dfrac{\partial \hat{v}}{\partial t} + \hat{u}\dfrac{\partial \bar{v}}{\partial x} = 0. \end{cases} \qquad (67)$$

The subsystem (67) can be rewritten as:

$$\dfrac{\partial \mathbf{W}}{\partial t} + \mathcal{B}(\mathbf{W})\partial_x \mathbf{W} = 0$$

$$\mathcal{B}(\mathbf{W}) = \begin{pmatrix} 0 & 0 & 0 & 0 & 0 \\ 0 & 0 & 0 & 0 & 0 \\ \dfrac{\hat{u}\hat{v}}{h} & 0 & 0 & \hat{v} & \hat{u} \\ 0 & 0 & 0 & 0 & 0 \\ 0 & 0 & \hat{u} & 0 & 0 \end{pmatrix}$$

Ones has five eigenvalues $\lambda_{1,2,3} = 0;\ \lambda_{4,5} = \pm\hat{u}$ and five eigenvectors that are

$$R_1 = \begin{pmatrix} 1 \\ 0 \\ 0 \\ 0 \\ -\dfrac{\hat{v}}{h} \end{pmatrix},\ R_2 = \begin{pmatrix} 0 \\ 1 \\ 0 \\ 0 \\ 0 \end{pmatrix},\ R_3 = \begin{pmatrix} 0 \\ 0 \\ 0 \\ 1 \\ 0 \end{pmatrix},\ R_4 = \begin{pmatrix} 0 \\ 0 \\ 1 \\ 0 \\ 1 \end{pmatrix},\ R_5 = \begin{pmatrix} 0 \\ 0 \\ 1 \\ 0 \\ -1 \end{pmatrix}.$$

We have $\det\left(R_1, R_2, R_3, R_4, R_5\right) = -2 \neq 0$. Then, the distortion waves subsystem (67) is hyperbolic.

**Theorem 3.1**

*Suppose the set of admissible solution by:*

$\mathcal{W} = \left\{\mathbf{W} \in \mathbb{R}^5,\ \hat{\mathbf{u}} > 0,\ h > 0\right\}$. *Then, the nonconservative system given by (59) is hyperbolic for all* $\mathbf{W} \notin \mathcal{W}$. *The eigenvalues of* $\mathcal{A}_\mathbf{n} \in \mathbb{R}^5$ *reads:*

$$\lambda_{1,5} = \mathbf{u}.\mathbf{n} \pm \sqrt{3\left(\hat{\mathbf{u}} \otimes \hat{\mathbf{u}}\right)\left(\mathbf{n} \otimes \mathbf{n}\right) + gh},\ \lambda_{2,4} = \mathbf{u}.\mathbf{n} - \hat{\mathbf{u}}.\mathbf{n},\quad \lambda_3 = \mathbf{u}.\mathbf{n}. \qquad (68)$$

*and we can always find a complete set of eigenvector for* $A_n \in \mathbb{R}^5$. *In addition, the* $\lambda_{1,5}$ − *fields are genuinely nonlinear, while* $\lambda_{2,3,4}$ − *field are linearly degenerate. A set of independent* $\lambda_i$ − *Riemann invariants,* $i = 1,...,5$ *can be found for each eigenvalues.*

**Theorem 3.2**

*Consider that the distortion disappear i.e.* $\hat{\mathbf{u}} = 0$. *Suppose* $h > 0$. *Then the nonconservative system obtained by removing the turbulence is still hyperbolic. Moreover, the eigenvalues of this model read* [9]*:*

$$\lambda_{1,3} = \mathbf{u}.\mathbf{n} \pm \sqrt{gh}, \quad \lambda_2 = \mathbf{u}.\mathbf{n}. \qquad (69)$$

*We can always find a complete set of eigenvector when* $A_n \in \mathbb{R}^3$. *In addition, the* $\lambda_{1,3}$ − *field are genuinely nonlinear, while* $\lambda_2$ − *field are linearly degenerate. A resonance condition, coalescence condition and a set of independent* $\lambda_i$ − *Riemann invariants can be found* [9]*.*

When the turbulence decreases over time, we have:

$$\lambda = \mathbf{u}.\mathbf{n} \pm \sqrt{3(\hat{\mathbf{u}} \otimes \hat{\mathbf{u}})(\mathbf{n} \otimes \mathbf{n}) + gh} \to \lambda = \mathbf{u}.\mathbf{n} \pm \sqrt{gh}.$$

Such context can appear after a shock or with the presence of a hydraulic jump. Subsequently, the effect of the distortion becomes negligible, resulting in a shallow water flow. In sediment transport context, the influence of the distortion wave has been well observed [14]. It has been demonstrated that this wave increases the wave front of the sediment concentration [14]. The following defines the Froude number for both first-order SWE of [1] and high-order SWE Eq. (36).

### 3.3 Rankine-Hugoniot relations for discontinuous solutions

The Rankine-Hugoniot relations describe the relations between the solution of the model in the unshocked medium and the shocked medium vary. However, these relations are only valid for discontinuities of negligible thickness (as showed in Fig. 2 below).

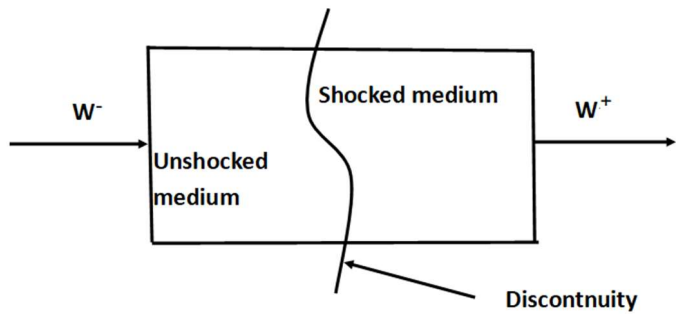

**Fig. 2:** *Relations between solutions of the model*

When discontinuities become strong, these relations cannot be applied.

- **The Rankine-Hugoniot (R-H) relations**

We established the R-H relations for the proposed model by assuming that the thickness of the discontinuity is negligible. Notice that $\bar{\mathbf{u}} \otimes \bar{\mathbf{u}} = (\bar{\mathbf{u}}.\mathbf{n})\bar{\mathbf{u}} = (\mathbf{u}_n)\bar{\mathbf{u}}$, the discontinuous solutions of our model verify the R-H relations follows:

$$\begin{cases} [[h(\mathbf{u}_n - D_n)]] = 0, \\ \left[\left[h\bar{\mathbf{u}}(\mathbf{u}_n - D_n) + \frac{gh^2}{2}\mathbf{n} + \overline{\rho h}(\hat{\mathbf{u}} \otimes \hat{\mathbf{u}})\mathbf{n}\right]\right] = 0, \\ \left[\left[h\bar{\mathbf{u}}(\mathbf{u}_n - D_n)\mathcal{E} + \mathbf{n}^T\left(\frac{gh^2}{2}\mathbf{I} + \overline{\rho h}(\hat{\mathbf{u}} \otimes \hat{\mathbf{u}})\right)\bar{\mathbf{u}}\right]\right] = 0, \end{cases} \qquad (70)$$

where $\mathbf{I}$ is the identity $2 \times 2$ matrix.

We have defined the jumps and average operators by $[[\bullet]] = (\bullet)^+ - (\bullet)^-$ and $\{\bullet\} = \frac{(\bullet)^+ + (\bullet)^-}{2}$ respectively; $(\bullet)^+, (\bullet)^-$ being the right and left limit of values at the discontinuity surface. In Eq. (70), $D_n$ is the normal of the surface, and $\mathbf{n}$ is the normal unit vector to the surface. We denote also by $\mathbf{s}$ the tangent unit vector to the surface such that $(\mathbf{n}, \mathbf{s})$ form a cartesian basis. Using $\mathbf{u} = \mathbf{u}_n \mathbf{n} + \mathbf{u}_s \mathbf{s}$, one obtains the following equation from the momentum equation:

$$\left[\left[h\bar{\mathbf{u}}_n(\bar{\mathbf{u}}_n - D_n) + \frac{gh^2}{2}\right]\right] = 0,$$

$$\left[\left[h(\bar{\mathbf{u}}_n - D_n)\bar{\mathbf{u}}_s + h\mathbf{s}^T(\hat{\mathbf{u}} \otimes \hat{\mathbf{u}})\mathbf{n}\right]\right] = 0,$$

and from the kinetic energy equation:

$$\left[\left[h\bar{\mathbf{u}}_n(\bar{\mathbf{u}}_n - D_n)\mathcal{E} + \bar{\mathbf{u}}_n\mathbf{n}^T\left(\frac{gh^2}{2}\mathbf{I} + h(\hat{\mathbf{u}} \otimes \hat{\mathbf{u}})\right)\mathbf{n}\right]\right] = 0,$$

$$\left[\left[\bar{\mathbf{u}}_s(\bar{\mathbf{u}}_n - D_n) + \bar{\mathbf{u}}_s h\mathbf{s}^T(\hat{\mathbf{u}} \otimes \hat{\mathbf{u}})\mathbf{n}\right]\right] = 0.$$

Often, we distinguish two types of discontinuities: contact discontinuities (interfaces) where $\mathbf{u}_n = D_n$ (the normal velocity is continuous across the contact wave), and shock discontinuities where $\mathbf{u}_n \neq D_n$.

- **Contact discontinuities**

Here, we consider the fact that $\mathbf{u_n} = D_n$. In this case the above momentum, kinetic energy are equivalent respectively to:

$$\left[\!\left[\frac{gh^2}{2} + h\mathbf{n}^T\left(\hat{\mathbf{u}} \otimes \hat{\mathbf{u}}\right)\mathbf{n}\right]\!\right] = 0,$$

$$\left[\!\left[h\mathbf{s}^T\left(\hat{\mathbf{u}} \otimes \hat{\mathbf{u}}\right)\mathbf{n}\right]\!\right] = 0.$$

and

$$\left[\!\left[\bar{\mathbf{u}}_\mathbf{n}\mathbf{n}^T\left(\frac{gh^2}{2}\mathbf{I} + h\left(\hat{\mathbf{u}} \otimes \hat{\mathbf{u}}\right)\right)\mathbf{n}\right]\!\right] = 0,$$

$$\left[\!\left[\bar{\mathbf{u}}_s h\mathbf{s}^T\left(\hat{\mathbf{u}} \otimes \hat{\mathbf{u}}\right)\mathbf{n}\right]\!\right] = 0.$$

Two types of contact discontinuities must be distinguished. The first type is identified by the vanishing of the tangential component of the stress vector on each side of the contact discontinuity. Thus a priori, we have $[\![\mathbf{u}_s]\!] \neq 0$. The second type of contact discontinuities is where the component of the positive velocity vector (or the acceleration vector) $\mathbf{s}^T\left(\hat{\mathbf{u}} \otimes \hat{\mathbf{u}}\right)\mathbf{n}$ is continuous, but sliding is not allowed.

- **Shock waves**

The kinematic wave equation describes the shock wave, which is an intrinsic feature of hyperbolic equations. We now consider that there is a discontinuity at the interface i.e. $\mathbf{u_n} \neq D_n$. The R-H relations (70) become:

$$\begin{cases} \left[\!\left[h(\bar{\mathbf{u}}_n - D_n)\right]\!\right] = 0, \\ \left[\!\left[h\bar{\mathbf{u}}_\mathbf{n}(\bar{\mathbf{u}}_n - D_n) + \frac{gh^2}{2} + h\mathbf{n}^T\left(\hat{\mathbf{u}} \otimes \hat{\mathbf{u}}\right)\mathbf{n}\right]\!\right] = 0, \\ \left[\!\left[h\bar{\mathbf{u}}_\mathbf{n}(\bar{\mathbf{u}}_n - D_\mathbf{n})\mathcal{E} + \bar{\mathbf{u}}_\mathbf{n}\mathbf{n}^T\left(\frac{gh^2}{2}\mathbf{I} + h\left(\hat{\mathbf{u}} \otimes \hat{\mathbf{u}}\right)\right)\mathbf{n}\right]\!\right] = 0, \\ \left[\!\left[h\bar{\mathbf{u}}_\mathbf{n}(\bar{\mathbf{u}}_n - D_\mathbf{n})\mathcal{E} + \bar{\mathbf{u}}_s h\mathbf{s}^T\left(\hat{\mathbf{u}} \otimes \hat{\mathbf{u}}\right)\mathbf{n}\right]\!\right] = 0, \end{cases} \quad (71)$$

Sliding along the shock surfaces is permitted in this case. However, the R-H relation for this system is inadequate to fully represent the structure of strong discontinuities. The nonconservative nature of sediment transport in equations for nonhomogeneous shear shallow water flows poses significant mathematical and numerical challenges.

**Remark**

*The Rankine-Hugoniot relations are valid only for weak shocks. When the shocks become too large, these relations become questionable and the numerical scheme based on them can fail [19]. These drawbacks are well observed in numerical tests and can guide the design of numerical methods to solve nonconservative equations with large shocks.*

### 3.4 Steady states solutions

In context of numerical simulations, the preservation of steady states (or stationary solutions) obtained by taking a vanishing time derivative in (59) or (39) remains of prime importance. An example of such stationary solutions are the well-known steady states of a lake at rest, which are obtained with zero water discharge and zero turbulence.

$$\overline{\mathbf{u}} = 0, \ \hat{\mathbf{u}} = 0, \ h = cste, \tag{72}$$

We consider relevant solution of the model obtained at the steady states; i.e. when $\dfrac{\partial \mathbf{W}}{\partial t} = 0$ in (59). We have:

$$\begin{cases} h\overline{\mathbf{u}} = cste, \\ \nabla \cdot \left( h\overline{\mathbf{u}} \otimes \overline{\mathbf{u}} + h\hat{\mathbf{u}} \otimes \hat{\mathbf{u}} + \dfrac{gh^2}{2} \right) = \mathbf{F}, \\ \nabla \cdot \left( \dfrac{h\overline{\mathbf{u}}}{2} \left( \overline{\mathbf{u}} \otimes \overline{\mathbf{u}} + 3\hat{\mathbf{u}} \otimes \hat{\mathbf{u}} \right) \right) + gh\overline{\mathbf{u}} \nabla h = \mathbf{F}\overline{\mathbf{u}}. \end{cases} \tag{73}$$

Obtaining steady states with non-zero discharge, also known as moving steady states, in two-dimensional space can be challenging. However, we present here 1D steady states. We begin by giving a more general friction term without the turbulence effect as follows:

$$\mathbf{F} = -K_{a,b} \dfrac{1}{h^b} |q|^a q, \quad a,b \in \mathbb{R}, \ K_{a,b} \in \mathbb{R}^*_+ \tag{74}$$

Using the friction source term given by (74) and from Eq., the 1D moving steady states read:

$$\begin{cases} h\overline{u} = q_0 \neq 0, \ , \ \dfrac{\hat{u}}{h} = \mathbb{P}_0, \ h > 0, \ u \neq 0, \\ \partial_x \left( \dfrac{q_0^2}{h} + h^3 \mathbb{P}_0^2 + \dfrac{gh^2}{2} \right) + K_{a,b} \dfrac{1}{h^b} |q_0|^a q_0 = 0, \end{cases} \tag{75}$$

with the above relation (75) we will design a set of moving steady state for regular $h$.

Let denote by $\mathcal{W}_I$ the space given by:

$$\mathcal{W}_I = \left\{ h \in \mathcal{W}, \ h \in C(I), \ h\overline{u} = q_0 \in \mathbb{R}^*, \ \dfrac{\hat{u}}{h} = \mathbb{P}_0 \in \mathbb{R}^+ \right\}, \tag{76}$$

where *I* is an interval of $\Omega \subset \mathbb{R}$.

For all $h \in \mathcal{W}_I$ the relation (75) becomes:

$$-\frac{1}{b-1}\frac{q_0^2}{h}\partial_x h^{b-1} + \frac{\mathbb{P}_0^2}{b+2}\partial_x h^{b+2} + \frac{gh}{b+2}\partial_x\left(h^{b+2}\right) + K_{a,b}\,|q_0|^a\,q_0 = 0 \qquad (77)$$

The steady states given by the system are highly nonlinear, and exact preservation remains an open problem.

## Conclusion

This work presents new contributions to the theory of geophysical turbulent flows with a free surface. Several new models in different form are presented, all based on new and relevant hydrodynamic and physical concepts that have not been explored previously. One of the main concepts studied is the additional velocity that arises when the flow becomes turbulent. It is well-known that the additional velocity arising from the turbulent nature of the fluid cannot be neglected and plays a role in the acceleration of the flow.

It has been observed that when the water flow becomes turbulent, the movement of the water fluid is influenced by the formation of vortices or eddies, as well as fluctuations. As the flow becomes turbulent, the fluctuations correlate and can therefore affect the mean velocity. This implies that fluid particle trajectories can be disturbed by turbulence. The wave structure of the introduced HSW model is noteworthy because it includes additional new waves to the familiar classical ones. Waves in a turbulent flow are faster than in an unsteady or steady flow.

The proposed work builds upon several relevant studies in shallow water flow and addresses modeling insufficiencies found in the literature. The results of this research will be relevant to hydraulic engineers and even coastal engineers due to the turbulent nature of water flow.

## Conflict of Interests

The author declares that there is no conflict of interest regarding the publication of this paper.

## Acknowledgment

The author would like to thank an anonymous referee for giving very helpful comments and suggestions that have greatly improved this paper.